# The effect of long-range forces on cold-atomic interaction: Ps-H system

Hasi Ray

Department of Science, National Institute of T. T. T. and Research Kolkata, Salt Lake City, Kolkata-700106, India
Email: hasi_ray@yahoo.com

**Abstract:** The s-wave elastic phase-shifts and s-wave elastic cross sections are studied to find the effect of long-range forces on cold-atomic interactions using a modified static-exchange model for Ps-H system. A Feshbach resonance in the triplet channel using the modified static-exchange model at the energy ~ $3 \times 10^{-6}$ eV caused by long-range forces in the Ps-H system is being reported.

**Introduction:**

The behavior of ultra-cold atomic systems and observation of Bose-Einstein condensation (BEC) at the laboratory have created a great adventure since the last two decades [1-3]. The most important basic question is: why a few of the systems are creating BECs and most are not? As for example, the experimental scientists tried hard to achieve BEC using spin polarized hydrogen beam [4], but failed to achieve it. On the otherhand, generally the alkali atoms are very useful to form a BEC.

In the ultra-cold systems close to BEC, the kinetic energies of the atoms are negligibly small and the interaction time is much larger than normal atomic interactions. The densities are ~ $10^{10}$ atoms/cm$^3$; the interatomic separation is ~ $10^{-4}$ to $10^{-3}$ cm, it is $10^4$ to $10^5$ times larger than atomic dimension ~ $10^{-8}$ cm. Two of the slowly moving atoms can come close to each other when all others are far apart. In this approximation, the atomic collision physics can provide reliable information about the cold-atomic systems. The most delicate interactions like exchange, van der Waals interaction etc. play a very important role in scattering theory at low energies [5-7] since due to the low kinetic energy, the projectile interacts with the target for a longer time. The electron-electron exchange is a short-range force, attractive in singlet- and repulsive in triplet- channels. The van der Waals force is due to the long-range dipole-dipole interaction between two atoms is attractive in nature. It seems that the latter is responsible for BEC formation due to the fact that the polarizability for alkali atoms are generally much larger than normal atomic hydrogen.

The van der Waals interaction between two atoms is defined as $V_{van}(R) = -\frac{C_W}{R^6}$, where $R$ is the interatomic separation and $C_W$ is the *van der Waals force constant* which depends on the polarizabilities of the two atoms and many other factors present in the system. As for example,

$C_W = 6.499026$ a.u. for H-H system [8];

$C_W = 69.56946$ a.u. for Ps-H system;

$C_W = 415.93177$ a.u. for Ps-Ps system [9].

It should be noted that the polarizability of H is 4.5 a.u. and the polarizabilty of Ps is eight times higher than normal H. When the time of electromagnetic radiation to propagate between atoms becomes comparable to the lifetime of the fluctuating dipoles, the correlation between the nearby dipoles are weakened and the attractive energy is reduced, ultimately decaying as ~ $R^{-7}$ instead of ~$R^{-6}$. This is known as retardation [10] and is defined as $V_{retd}(R) = -\frac{23 \hbar c \alpha_a \alpha_b}{4\pi \pi^7}$, where $\alpha_a, \alpha_b$ are the polarizabilities of the two atoms, $c$ is the velocity of light $=1/\alpha=137.036$ in a.u. with $\alpha$, the fine structure constant. A comparison of the magnitudes of van der Waals interaction ($V_{van}$) and retardation potential ($V_{retd}$) at different interatomic separations ($R$) for H-H, Ps-H and Ps-Ps systems is presented in Table 1. The values of the interatomic separations ($R'$) when both the interactions have equal magnitudes i.e. $V_{van}(R')=V_{retd}(R')$ are (a) $R'=584\,a_0$ for Ps-H system and (b) $R'=782\,a_0$ for H-H and Ps-Ps systems, when $a_0$ is the Bohr radius. It is noted that when $R$ is greater than $R'$, the magnitude of $V_{retd}$ is smaller than $V_{van}$ but when $R$ is less than $R'$, the magnitude of $V_{retd}$ is greater than $V_{van}$.

It is assumed that the long-range dipole-dipole interaction between two atoms e.g. van der Waals interaction causes the BEC formation. If so then the proper inclusion of the potential due to it, would create a Feshbach resonance near the critical temperature if BEC formation becomes possible in the system. The idea has motivated the present work to modify the static-exchange model [11] including the effect of long-range dipole-dipole interaction and to study the s-wave elastic phase shift and s-wave elastic cross section at cold energies. At the first step, positronium (Ps) and hydrogen (H) system is chosen; Ps is the lightest hydrogen like system and a good probe to investigate new physics. We study both the singlet (+) and triplet (-) channels. In the singlet channel, spins of two system electrons are antiparallel; so the short-range exchange forces between electrons are attractive and dictates the possibility of molecule formation. In the triplet channel, the electron spins are parallel and the short-range exchange forces between electrons are repulsive. In a scattering process, when the energy of the projectile is high it generally ignores all the delicate interactions like exchange and van der Waals interactions; but at low energies these interactions play an important role to reveal the fundamental physics of the system.

In the static-exchange model [11], the system wavefunction is written using a single-channel eigen-state expansion and the reaction channel should be the elastic channel. In this model, to solve Schrodinger wave equation the effect due to short-range exchange forces is considered into account but no long-range forces. So we modify the method adding the potential due to long-range dipole-dipole interaction with the static-exchange potential. In the present modified static-exchange (MSE) model, the first-Born term due to long-range dipole-dipole interaction is defined as

$$V_{add}=\int d\vec{r_1}\int d\vec{r_2}\int d\hat{R}\int_{2a_0}^{R'} dR R^2 \psi_f(\vec{R},\vec{r_1},\vec{r_2})\left\{-\frac{C_W}{R^6}\right\}\psi_i(\vec{R},\vec{r_1},\vec{r_2})+$$

$$\int d\vec{r_1}\int d\vec{r_2}\int d\hat{R}\int_{R'}^{\infty} dR R^2 \psi_f(\vec{R},\vec{r_1},\vec{r_2})\left\{-\frac{23\hbar c\alpha_a\alpha_b}{4\pi\pi^7}\right\}\psi_i(\vec{R},\vec{r_1},\vec{r_2})$$

Here $\psi_i(\vec{R},\vec{r_1},\vec{r_2})$ and $\psi_f(\vec{R},\vec{r_1},\vec{r_2})$ represent respectively, the initial and the final channel wavefunctions of the system. Since two atoms never can occupy the same space, the minimum values of the interatomic separation is chosen as equal to twice the Bohr radius.

The effective range theory predicts phase shifts ($\eta_0$) as a function of energy or momentum ($k$) of the projectile so that

$$k\cot\eta_0(k)=-\frac{1}{a}+\frac{1}{2}r_0 k^2$$

where $a$ is the scattering length and $r_0$ is the range of the potential. Hence, the scattering length is defined as, $a=-\lim_{k\to 0}[\frac{\tan\eta_0(k)}{k}]$. A positive scattering length indicates the possibility of binding and a negative scattering length indicates the possibility of no binding in the system. In Ps-H system, both the singlet and triplet scattering lengths are positive [12,13].

**Table 1**. Variation of the strengths of van der Waals interaction and retardation potential with the variation of interatomic separation ( R ):

| Interatomic separation $R$ (a.u.) | H-H system | | Ps-H system | | Ps-Ps system | |
|---|---|---|---|---|---|---|
| | $V_{van}(R)$ a.u. | $V_{retd}(R)$ a.u. | $V_{van}(R)$ a.u. | $V_{retd}(R)$ a.u. | $V_{van}(R)$ a.u. | $V_{retd}(R)$ a.u. |
| 50 | 4.16E-10 | 6.50E-09 | 4.45E-09 | 5.20E-08 | 2.66E-08 | 4.16E-07 |
| 100 | 6.50E-12 | 5.08E-11 | 6.96E-11 | 4.06E-10 | 4.16E-10 | 3.25E-09 |
| 300 | 8.91E-15 | 2.32E-14 | 9.54E-14 | 1.86E-13 | 5.70E-13 | 1.49E-12 |
| 500 | 4.16E-16 | 6.50E-16 | 4.45E-15 | 5.20E-15 | 2.66E-14 | 4.16E-14 |
| 584 | | | 1.75E-15 | 1.75E-15 | | |
| 600 | 1.39E-16 | 1.81E-16 | 1.49E-15 | 1.45E-15 | 8.91E-15 | 1.16E-14 |
| 782 | 2.84E-17 | 2.84E-17 | | | 1.82E-15 | 1.82E-15 |
| 1000 | 6.50E-18 | 5.08E-18 | 6.96E-17 | 4.06E-17 | 4.16E-16 | 3.25E-16 |
| 2000 | 1.08E-19 | 3.97E-20 | 1.08E-18 | 3.10E-19 | 6.49E-18 | 2.53E-18 |
| 3000 | 8.91E-21 | 2.32E-21 | 9.54E-20 | 1.86E-20 | 5.70E-19 | 1.49E-19 |
| 5000 | 4.16E-22 | 6.50E-23 | 4.45E-21 | 5.20E-22 | 2.66E-20 | 4.16E-21 |

**Theory:**

The total wavefunction of a system comprising Ps and H atoms [11] may be written as

$$\Psi^{\pm}(r_p, r_1, r_2) = \frac{1}{\sqrt{2}}(1 \pm P_{12}) \sum_{nv} \varphi_n(r_2) \eta_v(\rho_1) F_{nv}^{\pm}(R_1) \qquad \ldots(1)$$

with $\vec{\rho}_1 = \vec{r}_p - \vec{r}_1$ and $\vec{R}_1 = \frac{1}{2}(\vec{r}_p + \vec{r}_1)$. Here $\vec{r}_1$, $\vec{r}_2$ are the position vectors of the two electrons and $\vec{r}_p$ that of positron; $\varphi_n$ and $\eta_v$ represent respectively the wavefunctions of H and Ps; $P_{12}$ is the exchange operator. In writing the total wavefunction of the system, the spin components of the three particles are neglected. It is shown by Fraser [14] that explicit dependence of the spin function in writing the wavefunction is not required, and finally, the scattering parameters can be obtained by proper spin averaging. The total Hamiltonian of the system is

$$H = -\frac{1}{2}\nabla_{\vec{r}_p}^2 - \frac{1}{2}\nabla_{\vec{r}_1}^2 - \frac{1}{2}\nabla_{\vec{r}_2}^2 + \frac{1}{\vec{r}_p} - \frac{1}{\vec{r}_1} - \frac{1}{\vec{r}_2} - \frac{1}{|\vec{r}_p - \vec{r}_1|} - \frac{1}{|\vec{r}_p - \vec{r}_2|} + \frac{1}{|\vec{r}_1 - \vec{r}_2|} \qquad \ldots(2)$$

The Schrodinger equation for the system is

$$H\Psi^{\pm}(r_p, r_1, r_2) = E\Psi^{\pm}(r_p, r_1, r_2) \qquad \ldots(3)$$

The formally exact Lippman-Schwinger type coupled integral equations for transition amplitude in momentum space are given by Ghosh et al [15] :

$$\langle \vec{k}'n'v'|Y^{\pm}|\vec{k}nv\rangle = \langle \vec{k}'n'v'|B^{\pm}|\vec{k}nv\rangle + \sum_{n''v''}d\vec{k}''\frac{\langle \vec{k}'n'v'|B^{\pm}|\vec{k}''n''v''\rangle\langle \vec{k}''n''v''|Y^{\pm}|\vec{k}nv\rangle}{E-E''+i\varepsilon} \quad \ldots(4)$$

with $\quad \langle \vec{k}'n'v'|Y^{\pm}|\vec{k}nv\rangle = \langle \vec{k}'n'v'|Y_{11}|\vec{k}nv\rangle \pm \langle \vec{k}'n'v'|Y_{21}|\vec{k}nv\rangle \quad \ldots(5)$

and $\quad \langle \vec{k}'n'v'|B^{\pm}|\vec{k}nv\rangle = \langle \vec{k}'n'v'|B_{11}|\vec{k}nv\rangle \pm \langle \vec{k}'n'v'|B_{21}|\vec{k}nv\rangle \quad \ldots(6)$

where the unknown transition matrix elements $Y_{11}$ and $Y_{21}$, stand for the direct and exchange channels. Similarly $B_{11}$ and $B_{12}$ are the known first Born matrix elements for direct and exchange channels.

The static-exchange model permits only the elastic channel in the eigen state expansion of the system wavefunction. The elastic scattering is defined in the way that all the atoms should be in the same energy states in both the initial and final channels and $|\vec{k}_i|=|\vec{k}_f|$ if $\vec{k}_i$ and $\vec{k}_f$ are projectile momenta in the initial and final channels.

**Results and Discussion :**

We study the s-wave elastic phase shifts ($\eta_0^{\pm}$) and s-wave elastic cross sections. We consider both the atoms in the ground states. Our results for s-wave elastic phase shifts for the singlet (+) and triplet (-) channels using static-exchange model and the modified static-exchange model are presented in Figure 1. A sudden change in phase shifts by $\pi$ radian at the energy $\sim 3\times10^{-6}$ eV is observed in the triplet channel using the modified static-exchange model. It is an indication of the presence of a resonance. To ascertain the presence of resonance, we study the s-wave elastic triplet cross sections and plot them in Figure 2. A sharp peak in triplet cross section using modified static-exchange model confirmed the presence of a Feshbach resonance in the triplet channel at the energy $\sim 3\times10^{-6}$ eV. However, the singlet channel is almost insensitive to the long-range effects; it is in agreement with the earlier findings [5-7]. It should be noted that 1 eV energy is equivalent to a temperature $\sim 10^4$ degree Kelvin. It is evident that the long-range attractive van der Waals force due to dipole-dipole interaction is responsible for occurring the Feshbach resonance.

**Conclusion** :

We report a Feshbach resonance in triplet channel in the cold energy region in Ps and H scattering when both the atoms are treated at ground states using a modified static-exchange model. In this model, we include the long-range effect e.g. the van der Waals interaction into the static-exchange model [11]. Using static-exchange model, no resonance is found in the cold-energy region. So it is evident that the attractive long-range van der Waals interaction is responsible for occurring the Feshbach resonance at the cold energy region in Ps and H system when both the atoms are at ground states with parallel electron spins, indicating the possibility of BEC formation.

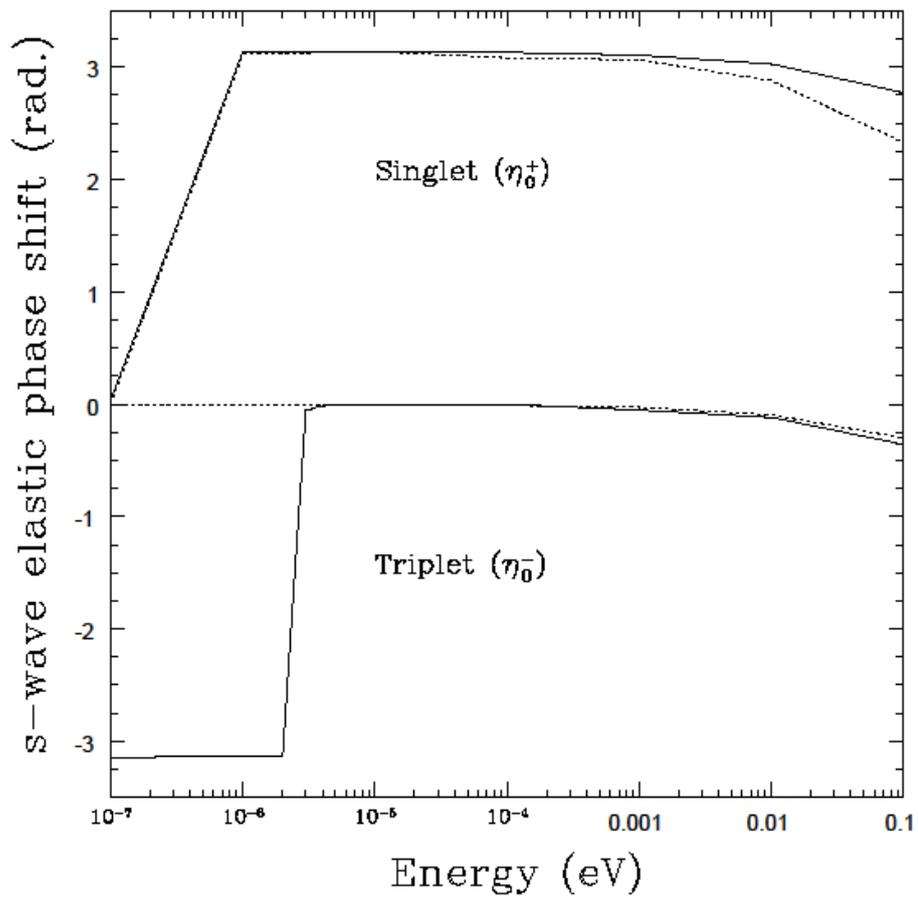

**Figure 1.** The s-wave elastic phase shifts in radian for the singlet (+) channel (upper curves) and triplet (-) channel (lower curves) using the static-exchange model (the dotted lines) and the modified static-exchange model (the solid lines) in the cold energy region.

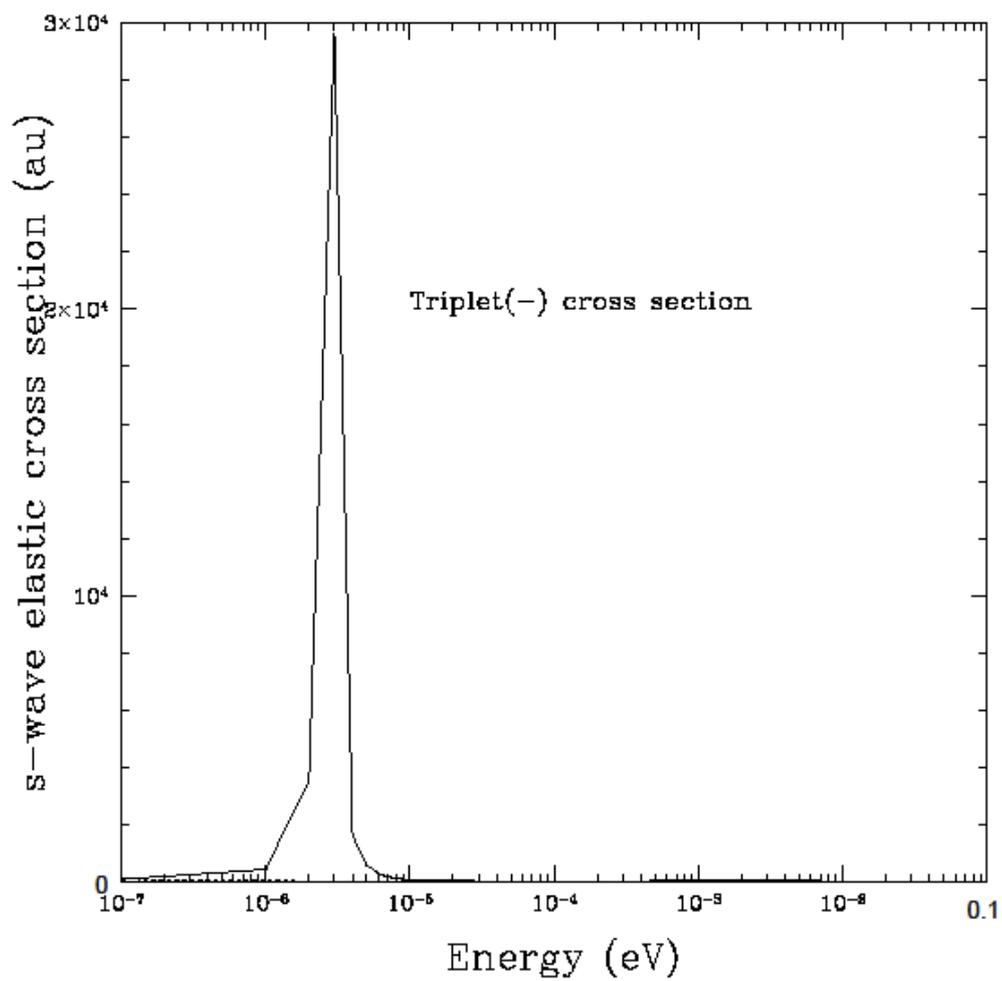

**Figure 2.** The s-wave elastic cross sections in a.u. for the triplet (-) channel using the static-exchange model (the dotted lines) and the modified static-exchange model (the solid lines) in the cold energy region.

**Acknowledgement**

The author acknowledges the financial support by SERC, DST Govt. of India through grant no. SR/WOS-A/PS-13/2009 dated 30.11.2009. She is thankful to Prof. S. K. Bhattacharyya, Director, NITTTR-K for providing the scope to implement her research project at the present institute. She is thankful to Prof. A. De, Head of Science Department, NITTTR-K for being the mentor. She is thankful to Prof. C. S. Unnikrishnan, TIFR, Mumbai for the academic help during her visit at TIFR, Mumbai on August 2008.